# The Dynamics of Trust: A Stochastic Lévy Model Capturing Sudden Behavioral Jumps

Mohamadali Berahman [a] and Madjid Eshaghi Gordji[b]*

[a]*Department of Mathematics, Faculty of science, Semnan University, Semnan, Iran*
[b]*Department of Mathematics, Faculty of science, Semnan University, Semnan, Iran*

*meshaghi@semnan.ac.ir

# The Dynamics of Trust: A Stochastic Lévy Model Capturing Sudden Behavioral Jumps

Trust is the invisible glue that holds together the fabric of societies, economic systems, and political institutions. Yet, its dynamics-especially in real-world settings remain unpredictable and difficult to control. While classical trust game models largely rely on discrete frameworks with limited noise, they fall short in capturing sudden behavioral shifts, extreme volatility, or abrupt breakdowns in cooperation.Here, we propose-for the first time a comprehensive stochastic model of trust based on Lévy processes that integrates three fundamental components: Brownian motion (representing everyday fluctuations), Poissonian jump intensity (capturing the frequency of shocks), and random distributions for jump magnitudes. This framework surpasses conventional models by enabling simulations of phenomena such as "sudden trust collapse," "chaotic volatility," and "nonlinear recoveries" dynamics often neglected in both theoretical and empirical studies.By implementing four key simulation scenarios and conducting a detailed parameter sensitivity analysis via 3D and contour plots, we demonstrate that the proposed model is not only mathematically more advanced, but also offers a more realistic representation of human dynamics compared to previous approaches. Beyond its technical contributions, this study outlines a conceptual framework for understanding fragile, jump-driven behaviors in social, economic, and geopolitical systems-where trust is not merely a psychological construct, but an inherently unstable and stochastic variable best captured through Lévy based modeling.

**Introduction**

Trust is one of the most fundamental components of human social behavior, playing a central role in the formation and stability of economic, political, and even cognitive relationships. At its simplest, trust can be described as a psychological state in which one individual (the trustor) allocates resources or privileges to another (the trustee) without any guarantee of reciprocation or return. Because this act is inherently accompanied by risk and uncertainty, analyzing its dynamics requires tools that go beyond classical psychology namely, game theory and advanced stochastic modeling.

One of the most widely used empirical frameworks for studying trust is the trust game introduced by Berg et al. (1995) [1]-a two-player setup in which the first player transfers an amount of capital to the second. This transfer signals an initial level of trust, while the second player has the discretion to either return a portion of the capital or keep it an action reflecting their level of trustworthiness. In a more rigorous theoretical formulation, Dasgupta (2000) [2] conceptualizes trust not as an expression of goodwill, but rather as a rational expectation about the behavior of others under conditions where direct monitoring is not possible. From this perspective, the trust game becomes a powerful tool for analyzing decision-making in environments characterized by incomplete information and behavioral uncertainty. Extending this theoretical line, the inequality-aversion model proposed by Fehr and Schmidt (1999) [3] suggests that fairness related preferences and sensitivity to inequality can promote cooperative behavior within the trust game, even in the absence of formal monitoring mechanisms. By integrating empirical evidence from competitive markets and voluntary interactions, their model provides an analytical framework for understanding ethics driven decisions in contexts of risk and opportunism. As part of broader theoretical efforts to explain the dynamics of trust, Nowak (2006)[4] identifies five key mechanisms in the evolution of cooperation: kin selection, direct reciprocity, indirect reciprocity, network structure, and group selection. These mechanisms help bridge the gap between individual and collective interests and account for the persistence of cooperative behaviors in populations. Among them, direct reciprocity aligns most closely with the logic of the trust game, where cooperation is sustained by the expectation of future return.

In a similar vein, (Axelrod & Hamilton,1981)[5], focusing on the iterated prisoner's dilemma, illustrate how repeated interaction, behavioral feedback, and mediating factors such as reputation, social networks, learning, and even emotions like envy or misunderstanding can significantly influence the emergence and stability of cooperation. In his earlier seminal work The Evolution of Cooperation (Axelrod, 1984) [6], he introduced a theoretical model for stable cooperation among self-interested agents. Strategies such as tit-for-tat, grounded in learning and rational retaliation, were presented as central drivers of sustained trust in repeated interactions concepts that directly underlie many subsequent analyses of the trust game.

Recent empirical studies provide further evidence for this theoretical linkage. For instance, Lo Iacono et al. (2024) [7] demonstrate that in societies characterized by high levels of social trust, there is a greater propensity to invest in collective solutions. Conversely, in low trust environments, individuals tend to favor isolated strategies and engage in suboptimal resource allocation. These findings once again underscore the pivotal role of trust as a foundation for sustainable cooperation and efficient distribution of resources.

In recent years, growing efforts have been devoted to more precisely analyzing individual motivations within the trust game and developing computational frameworks to simulate its dynamics. Espin et al. (2016)[8], by examining the dual roles of participants, revealed that trust and cooperation behaviors emerge from the interplay of multiple motives such as altruism, strategic self interest, fairness concerns, and even spite.

Taking a further step, Meylahn et al. (2025)[9] introduced a stochastic compartmental model based on social segmentation, dividing the population into three distinct groups-trusters, skeptics, and distrusters-and analyzing their interactions under the paradigm of bounded confidence. Their findings reveal that variables such as life expectancy and population size can drive the spread of distrust. Advanced methodologies such as fluid approximations, diffusion dynamics, and the Gillespie simulation algorithm have proven effective in capturing the temporal and structural complexities of trust more accurately than traditional deterministic models.

In the domain of agent based simulation, the PISKaS framework developed by Perez Acle et al. (2018)[10] models trust as a dynamic, nonlinear phenomenon. By circumventing the limitations of differential equation based models, this framework enables the exploration of micro level social behaviors and individual interactions, opening new avenues for empirical research on trust.

In another line of inquiry, Zeng (2010)[11] modeled trust within the context of competition among insurance firms, using stochastic differential games driven by Brownian motion to derive Nash equilibria in a zero sum setting. This approach offers a compelling blueprint for analyzing strategic decision making in high-risk, uncertain environments.

Along similar lines, Exarchos et al. (2019)[12] developed a computationally efficient framework for solving two player games with conflicting interests, leveraging a reformulation of the Hamilton Jacobi Isaacs equations alongside the Feynman Kac lemma. This model is particularly well suited for studying dynamic interactions in trust-related games, especially under conditions of uncertainty.

Finally, Fehr (2009) [13] defines trust behaviorally as a voluntary act of resource transfer without legal obligation, distinguishing it from mere risk taking and linking it closely to social preferences and economic beliefs. This perspective facilitates understanding cultural and national differences in trust levels and highlights the intercultural dimension of trust analysis.

Building on these efforts, Bornhorst et al. (2004)[14] conducted an empirical study using a repeated trust game, demonstrating that trust and trustworthiness behaviors emerge from a complex interplay of factors, including reinforcement learning, reciprocal interactions, and instrumental rationality. Their findings emphasize the crucial role of repeated interactions and the multi stage structure of the game in fostering and sustaining trust among participants.

On another front, Dasgupta (2010) [15], focusing on the concept of social capital, presents trust as a foundation for effective cooperation. He argues that well-managed social networks can strengthen public trust and enhance macro level productivity, whereas mismanagement of this capital may lead to institutional decay and economic regression.

At a more structural level, Kumar et al. (2020)[16] examined the trust game across various social network types complete, random, well mixed, and scale free networks and found that network topology generally has limited influence on the evolution of trust. Only under specific conditions, such as scale free networks with non normalized dynamics, was evidence found for the stable emergence of trust or trustworthiness. These results suggest that trust evolution is context-sensitive and cannot be solely attributed to topological network features.

Continuing the empirical and theoretical analyses, Charness et al. (2011)[17] demonstrated that knowledge of a trustee's past trustworthy behavior can enhance trust levels as effectively as classical reputation systems. Their findings suggest that even when financial incentives are minimal, investing in building a reputation as a reliable individual through mechanisms such as indirect reciprocity strengthens trust promoting behaviors.

In a critical review, Alós Ferrer and Farolfi (2019) [18] reassess the trust game as a primary tool for measuring individual differences in trust. By comparing the trust game with psychometric questionnaires and neuroimaging methods, they emphasize that a comprehensive explanation of trust-related behaviors requires integrating biological and behavioral dimensions connections that the trust game alone cannot fully reveal.

From an experimental design perspective, Cox (2007)[19] highlights the importance of distinguishing between social behavior components such as altruism, reciprocal interaction, trust, and fear of betrayal. He argues that separating models based on intention from those independent of the counterpart's intentions is essential for accurately analyzing trust driven behaviors.

At the theoretical level, Marsili and Zhang (1998)[20] introduced stochastic dynamics into game theory, demonstrating that random deviations from the assumption of perfect rationality can destabilize Nash equilibria in large populations. Utilizing a physics-inspired approach based on free energy functions, they quantitatively analyzed how such fluctuations contribute to trust instability in competitive environments.

In a cross cultural study, Ashraf et al. (2006)[21] combined investment and dictator games across three countries and found that expectations of return on investment and unconditional kindness are key drivers of initial trust, whereas trustworthiness is predominantly influenced by unconditional kindness. Their comparative analysis revealed that despite cultural differences, fundamental behavioral patterns related to trust exhibit universal commonalities.

Rooting trust in foundational theory, Axelrod and Hamilton (1981) [5] introduced the tit for tat strategy in the iterated prisoner's dilemma, symbolically showing that stable cooperation among self-interested actors can emerge without central authorities, relying solely on repeated interaction and reciprocal responses. This pattern implicitly suggests that even minimal initial trust is often a necessary condition for the emergence of cooperation.

Over the past few decades, the trust game has become a central framework for analyzing interpersonal relationships, social institutions, and economic responses. Numerous studies have sought to examine this complex phenomenon from diverse perspectives. For example, Charness and Dufwenberg (2006)[22] demonstrated that verbal promises can enhance trustworthy behavior, as individuals tend to avoid the guilt associated with violating social expectations. Similarly, the study by Croson and Buchan (1999)[23] investigated the roles of cultural and gender differences in trust responsiveness, revealing that women, acting as proposers, tend to offer higher returns.

Studies such as Kosfeld et al. (2005)[24] have explored the chemical underpinnings of trust, particularly the role of oxytocin in enhancing trusting behavior. Their findings indicate that oxytocin increases individuals' willingness to accept social risks, even when their overall risk tolerance remains unchanged. In the same vein, Bohnet and Zeckhauser (2004) [25] argue that the decision to trust is not merely a standard risk-taking behavior but involves a psychological cost arising from the possibility of betrayal.

Moreover, investigations into individual differences and psychometric dimensions of the trust game have revealed certain limitations. Specifically, Alós Ferrer and Farolfi (2019) [18] note that despite its experimental advantages, the trust game alone cannot fully capture the insights provided by neuroscience and psychometric data, highlighting the need for complementary tools such as neuroimaging or chemical stimulation.

Our proposed model, grounded in non Gaussian and jump driven Lévy processes, builds on these foundational critiques to introduce an innovative framework capable of precisely modeling sudden shocks whether neurological, cognitive, or contextual that result in sharp fluctuations in trusting or trustworthy behavior. In contrast to traditional models that largely treat trust as evolving in a continuous and smooth space, the Lévy framework, with its heavy tailed jump dynamics, allows for the incorporation of abrupt cognitive shocks such as neurochemical triggers, sudden revelations, or unexpected betrayals. Furthermore, in line with Alós Ferrer's recommendation to move beyond standard trust game paradigms, our analysis integrates neural data and supports multiscale and multitemporal modeling of trust behavior.

In pursuit of a more structured analysis of trust dynamics, various dynamic models have been developed. Lim (2020) [26] demonstrated that even in the absence of auxiliary mechanisms such as punishment or reputation systems, high levels of trust and trustworthiness can emerge in populations with asymmetric demographic parameters including differences in group sizes. These findings contrast with the predictions of classical evolutionary game theory, which often deems the stability of trust and trustworthiness unlikely. They suggest that under stochastic conditions and when demographic asymmetries are taken into account, promising and divergent outcomes may arise.

While Lim's study targets a critical gap between theoretical models and empirical observations, the scope of stochasticity remains confined to a relatively simple Markovian process with rare mutations (weak mutation). In contrast, our proposed model, by leveraging the more powerful Lévy process, offers a richer and more realistic structure for capturing sudden fluctuations and behavioral jumps among agents. This provides greater capacity to represent the cognitive and behavioral complexities that define real-world trust interactions.

For example, Zheng et al. (2024) [27],, in their attempt to explain the emergence of trust and trustworthiness in the Trust Game, employed a reinforcement learning algorithm known as Q-learning. They demonstrated that when agents make decisions based on past experiences and expectations of future returns, high levels of trust can emerge endogenously. In their model, the agent's experiential memory plays a fundamental role, captured through the accumulation of Q-values over time.

Our central idea in this study is inspired by this very notion of "accumulated experience" as a form of behavioral memory. However, we depart from the stepwise, Markovian framework of Q-learning and instead model memory using a Lévy process a stochastic process characterized by long range dependence, where the weight of past events decays more slowly. Unlike the Q-learning approach, which updates memory incrementally and discretely, the Lévy based model also allows for sudden jumps in trust levels-a phenomenon frequently observed in real human behavior.

Thus, we reinterpret the core insight of Zheng et al. [27] regarding the role of memory and reward accumulation, embedding it within a continuous-time stochastic framework that captures both long-term memory and discontinuous behavioral shifts.

On the other hand, kumar et al. (2020) [16] simulated the trust game across various network structures including fully connected, random, scale-free, and well mixed networks-to investigate the emergence of trust and trustworthiness. Their analysis revealed that, contrary to common assumptions, the structure of the network often has no significant impact on the emergence of trust. However, in scale-free networks with specific dynamics, they observed conditions under which trust arises endogenously, without necessarily being accompanied by a corresponding level of trustworthiness.

A key insight from this study that we incorporate into our model is the notion of fragility and imbalance in the simultaneous emergence of trust and trustworthiness within particular networked settings. In our framework, this behavioral asymmetry is captured via asymmetric Lévy processes, meaning that large jumps in the level of trust may occur without necessarily being mirrored in trustworthiness behavior. Moreover, by incorporating long-term memory through Lévy jumps, our model is able to reproduce the nuanced and complex dynamics observed by kumar et al. [16] through a continuous and memory based mechanism.

Therefore, their analysis provided a crucial insight: Lévy based modeling of trust must treat the dynamics of the trustor and the trustee as distinct and potentially unequal processes.

Other studies have emphasized the role of social and biological dimensions in the formation of trust. For instance, Charness, Du, and Yang (2011)[17], in an economic experiment, investigated the role of reputation in reinforcing trust and trustworthiness. Their key findings revealed that not only the past performance of an individual as a trustee but also their previous behavior as a trustor can serve as a reliable indicator of reputation and credibility. Intriguingly, individuals continued to engage in trust-building investments even when the immediate returns of trusting actions were low. This phenomenon can be explained through the mechanism of indirect reciprocity, whereby the observation of positive behavior by a third party leads to rewarding responses toward the original actor.

In our proposed model based on Lévy jump processes, this finding is explicitly represented. The Lévy framework enables the modeling of sudden and asymmetric jumps in individuals' reputations-whether in the role of a trustor or a trustee. Notably, the effort to build a reputation for trustworthiness, even in the absence of immediate returns, is captured in our model as a gradual increase in the probability of positive jumps in trust received from others.

Thus, indirect reciprocity is modeled through concepts such as non-local memory and unexpected large jumps, facilitated by Lévy distributions within the reputation structure. This allows past behaviors in both trustor and trustee roles to exert long-term influence on the accrual of future trust.

Glaeser et al. (2000) [28], by combining two behavioral experiments and a survey study, attempted to empirically measure two key components of social capital trust and trustworthiness. Their findings revealed that standard attitudinal survey questions have greater predictive power for trustworthy behaviors than for trusting behaviors. Conversely, trustor behavior in laboratory games is more strongly influenced by individuals' prior real world interaction experiences. The study also emphasized that social proximity between individuals increases both trust and trustworthiness, while racial or national differences tend to reduce trustworthiness.

The significance of this research lies in highlighting the role of past experiential memory and social factors in shaping trust behaviour an issue that gains increasing importance in dynamic trust modeling frameworks based on history dependent processes, such as Lévy processes.

In the following sections of this introduction, we will review other relevant studies in greater detail and subsequently elaborate on the innovative position of our proposed model based on the Lévy process.

In pursuit of a deeper understanding of trust and trustworthiness behaviors, numerous empirical and theoretical studies have sought to enrich our knowledge of this phenomenon. For instance, Johnson and Mislin (2011) [29], by compiling data from 162 iterations of the investment game developed by Berg et al. [1] (the Trust Game), involving over 23,000 participants, examined the effects of experimental design and geographical variation on levels of trust and trustworthiness. Their results demonstrated that variables such as payment randomness, partner simulation, experimenter-determined return rates, participants' dual roles, and student status significantly influence trust and trustworthiness behaviors. Furthermore, strong evidence of regional differences was reported, with trust levels in games conducted in Africa being significantly lower than those in North America.

These findings underscore the profound dependence of trust behaviors on social and environmental contexts an aspect often underrepresented in purely discrete or static models.

In contrast, our proposed framework, grounded in Lévy processes, enables the modeling of environmental and contextual dependence through jump driven parameters an approach that can capture the complexities of trust behavior with greater accuracy and realism.

In one of the most recent laboratory studies, Hofmeyr et al. (2023)[30] demonstrated that providing participants with precise and transparent information about the financial consequences of their decisions significantly increases both returns and willingness to send trust. Notably, the graphical presentation of the counterpart's social history enhanced the propensity to trust-highlighting the importance of "transparency" and "social memory" as two key variables in trust dynamics.

These factors are incorporated in our model through the non-local and memory bearing structure of the Lévy process, whereby informational shocks or sudden betrayals can transform trust behavior via large jumps.

From the perspective of structural development in the trust game, Abbass et al. (2015)[31] examined collective trust dynamics by proposing an N player version of the game. They found that even the presence of a small number of untrustworthy individuals can lead to a complete collapse of trust a phenomenon highly sensitive to initial conditions. This feature is simulated in our model through the initial parameters of the Lévy path, whereby a small initial shock can dramatically alter the entire trajectory of the trust system.

Furthermore, Boero et al. (2010)[32], using simulations based on dynamic networks, investigated the role of partner selection in trust formation. Their findings revealed that social networks lead to trust only when free riders are isolated and cooperators strengthen their ties. Our Lévy process based model, with its ability to represent sudden changes in trust levels, can effectively model natural punitive and rewarding mechanisms within network structures.

In the realm of behavioral preferences, Attanasi et al. (2019)[33] demonstrated that guilt aversion and belief-dependent preferences play fundamental roles in actors' responses within the trust game. Through a controlled experimental design, they examined the impact of initial information about the responder's reputation. In our proposed model, this phenomenon is represented via changes in the probabilities of positive jumps (resulting from favorable reputation) or negative jumps (due to initial distrust).

In a different approach, Chica et al. (2019)[34] modeled trust within the framework of the sharing economy and demonstrated that the presence of insurance and punitive mechanisms can accelerate the stabilization of trust among various actors. Their key finding revealed that the intensity of punishment or the extent of insurance coverage can sometimes have a counterproductive effect on trust. This behavioral duality is reproducible in our model through the heavy tailed distributions inherent in Lévy jumps, whereby policies may induce sudden and nonlinear effects on trust behavior.

Finally, given the limitations of traditional models such as Brownian motion (Wiener processes) or Markov decision processes, recent literature has increasingly emphasized the importance of Lévy processes. As noted by Bertoin (1996)[35] and Applebaum (2009)[36], Lévy processes combining continuous components (e.g., Brownian motion) and jump components (e.g., Poisson processes) possess the capability to model behaviors that are both gradual and discontinuous. Moreover, Cohen and Solan (2013) [37] explored the applications of Lévy processes in modeling stochastic decision making, while Sørensen, and Benth (2013)[38] proposed methods for simulating the trajectories of these processes.

In this study, for the first time, we model the trust game using the full structure of the Lévy process, incorporating jump parameters ($\lambda$, $\mu_J$), volatility ($\sigma$), and behavioral drift ($\gamma$). This framework enables the representation of trust dynamics within unstable, uncertain, and shock sensitive environments. The proposed model is capable of capturing sudden collapses of trust, gradual recoveries, and severe jumps arising from memories, betrayals, or new information. This capability is especially critical for analyzing trust related policy making at macro social, economic, and geopolitical levels.

This research leverages stochastic process theory, particularly Lévy processes, to propose a novel framework for modeling trust dynamics in social interactions. Lévy processes, as a generalization of Brownian motion, possess the capability to represent sudden jumps and heavy tailed distributions; hence, they substantially outperform classical Markovian models which primarily rely on assumptions of continuous and homogeneous changes in analyzing non gradual shifts, psychological shocks, and radical fluctuations in trust levels (Applebaum, 2009)[36]. Our proposed model, by more accurately reproducing the empirically observed dynamics in trust games, enables simulation of diverse behaviors such as sudden trust collapses, intense volatility, and nonlinear recoveries, going beyond traditional frameworks.

Previous literature has predominantly focused on stochastic processes with continuous time steps like Wiener processes or Markov models with Gaussian noise. For instance, Lim (2020)[26], employing stochastic evolutionary dynamics in asymmetric populations, demonstrated that trust and trustworthiness can persist without auxiliary mechanisms such as punishment or reputation. However, these models assume gradual, uniform behavioral changes accompanied by homogeneous noise, an assumption inadequate for capturing psychological shocks, sudden betrayals, or damaging information that undermine trust.

To overcome this limitation, our proposed framework models agent behavior in trust games using Lévy processes, which allow the representation of discrete, large, and asymmetric jumps. This approach not only enriches the mathematical structure of the dynamic model but also provides the ability to more accurately describe sudden fluctuations and more complex behaviors observed in both one shot and repeated trust games (Applebaum, 2009)[36]. Thus, our model enables a deeper and more realistic examination of trust dynamics in social, economic, and even political environments.

Previous studies indicate that despite numerous investigations in economics, psychology, and neuroscience analyzing trust, there remains a lack of mathematical models that simultaneously integrate long term memory, behavioral jumps, and responses to sudden shocks within a unified framework. Our proposed model fills this gap by leveraging Lévy processes. Unlike discrete approaches such as Q-learning, this model incorporates behavioral memory and experiential effects in a continuous distribution with heavy tails, allowing for large jumps and persistent effects of prior behaviors. Furthermore, positive and negative jumps are designed asymmetrically to reflect behavioral differences between trustor actions and responder responses, which are clearly observed in empirical data.

Overall, this study presents a model based on non Gaussian Lévy processes, enabling a more precise representation of the complex dynamics of trust and betrayal in experimental games. It provides a powerful conceptual and mathematical tool for analyzing discontinuous and multifaceted social phenomena such as distrust, psychological shocks, and reputation formation. This framework constitutes an innovative combination of game theory, behavioral dynamics, and stochastic mathematics, opening new horizons in cognitive sciences, social sciences, and behavioral economics research.

A review of prior studies reveals that each has focused on specific aspects of the complex puzzle of trust such as social behavior, network dynamics, intrinsic motivations, or risk and uncertainty yet none have comprehensively modeled this phenomenon while incorporating stochastic and jump dynamics. Our proposed framework, for the first time in the game theory literature, integrates continuous fluctuations, sudden jumps, and behavioral drifts to model the trust game.

This model not only explains sudden collapses and gradual reconstructions of trust but also facilitates the analysis of behavioral sensitivities, prediction of nonlinear changes, and the design of social control policies to reinforce trust in real-world systems. Numerical simulations including contour plots, stochastic trajectories, and cumulative mean analyses have demonstrated the model's effectiveness in complex scenarios and unstable environments.

Thus, this research, by presenting a novel framework and departing from traditional models, proposes new pathways for studying the dynamics of human trust across the fields of game theory, behavioral sciences, and social policy.

**materials and methods**

*The Trust Game*

In the Trust Game, first introduced by berg et al. (1995) [1], two players engage in a sequential interaction assuming the roles of trustor and trustee. The trustor is given an initial endowment (e.g., $10) and decides how much of it to send to the trustee. The

amount sent is then multiplied (typically tripled), and the resulting sum is received by the trustee, who subsequently decides how much of it to return to the trustor. Under classical equilibrium assumptions, it is predicted that the trustee will act in pure self interest and return nothing, anticipating which the trustor would choose not to send any amount at all. However, empirical findings have consistently shown that individuals tend to exhibit trust and reciprocity, highlighting the central role of these behaviors in shaping economic and social interactions (berg et al., 1995)[1].

The mathematical formulation of the trust game, following Lim (2020)[26], is as follows: In this game, the trustor begins with an initial endowment of one unit and transfers a portion of it to the trustee. The transferred amount is then multiplied by a factor and received by the trustee. Subsequently, the trustee returns a fraction of this amount back to the trustor. The payoffs for each player are defined as follows:

$$payoff_{investor} = 1 - p + pbr \tag{1}$$

$$payoff_{trustee} = pb(1 - r) \tag{2}$$

Additionally, the fitness function of each player, considering the selection intensity, is defined as:

$$f = 1 + \beta\pi \tag{3}$$

where denotes the player's average payoff (Lim, 2020)[26].

*Modeling Stochastic Dynamics with Jumps Using Lévy Processes*

For a formal definition, we adopt the framework presented in Applebaum (2009)[36], in which Lévy processes are rigorously analyzed through the lens of probability theory and stochastic calculus. To more accurately model discontinuous dynamics and rare but high impact events in the trust game, we employ Lévy processes. A Lévy process is one of the most fundamental classes of increasing stochastic processes, comprising a combination of Gaussian (continuous) noise and random (discontinuous) jumps. These processes enable the modeling of phenomena characterized by anomalous behavior, abrupt shifts, or explosive variance features commonly observed in social, economic, and psychological contexts, including sudden changes in trust between agents.

More precisely, a stochastic process $(X_t)_{t \geq 0}$, defined on a probability space $(\Omega, F, P)$, is called a Lévy process if it satisfies the following properties:

(1) Initial condition:

$X_0 = 0$, That is, the process starts at zero almost surely.

(2) Independent increments:

For any selection of times $0 \leq t_0 < t_1 < \cdots$, the random variables $X_{t_1} - X_{t_0}, \ldots, X_{T_n} - X_{t_{n-1}}$, are mutually independent.

(3) Stationary increments:

For all $s, t \geq 0$, the distribution of the increment $X_{t+s} - X_s$ depends only on the length of the interval t, and not on the starting point s.

(4) Cadlag paths:

With probability one, the sample paths of the process are right-continuous with left limits (i.e., they are cadlag functions of time).

Every Lévy process can be uniquely characterized through the Lévy–Khintchine theorem. According to this theorem, the characteristic function of a Lévy process $(X_t)_{t \geq 0}$ is given by:

$$E\left[e^{iuX_t}\right] = \exp\left\{t\left(iu\gamma - \frac{1}{2}u^2\sigma^2 + \int_{\mathbb{R}\setminus} (e^{iux} - 1 - iux\mathbf{1}_{|x|<1})\nu(dx)\right)\right\} \quad (4)$$

where:

- $\gamma \in \mathbb{R}$ is the drift term ;

- $\sigma^2 \geq 0$ is the variance of the Gaussian (Brownian) component;
- $\nu$ is the Lévy measure, which governs the intensity and distribution of jumps.

Here, $u \in \mathbb{R}$ represents the Fourier transform variable, while the three terms within the parentheses correspond respectively to the drift component, the Gaussian variance, and the cumulative effect of jumps. Together, these elements $(\gamma, \sigma^2, \nu)$ form the Lévy triplet, which uniquely characterizes the statistical properties of the Lévy process.

While the Wiener process models only continuous noise, and the Poisson process is limited to discrete jumps, Lévy processes are capable of capturing both types of dynamics within a unified framework. This makes them particularly well suited for modeling phenomena such as unstable trust, sudden collapses in confidence, or unexpected recoveries of trust in agent-based interactions. In this paper, we adopt this framework to define the dynamics of the Trust Game, wherein strategies or trust levels evolve over time as stochastic processes influenced by Lévy noise (Applebaum, 2009)[36].

In this study, we employ Lévy processes to model the dynamic behavior of trust over time, capturing both continuous and discontinuous components of evolution. These processes admit two mathematically equivalent yet conceptually distinct representations:

First, the characteristic function representation, which is uniquely determined by the Lévy–Khintchine theorem. This formulation specifies the statistical structure of the process and is described by the Lévy triplet $(\gamma, \sigma^2, \nu)$, where $\gamma$ denotes the drift, $\sigma^2$ the Gaussian variance, and $\nu$ the Lévy measure governing the intensity and size distribution of jumps. This representation provides a powerful analytical tool for studying key probabilistic properties of the process, such as variance, expected value, and the likelihood of abrupt changes.

Second, the stochastic integral representation is particularly well-suited for numerical simulations of the process path. In this formulation, the trust process is modeled as the sum of three fundamental components: a drift term (representing the baseline trend of trust), a Gaussian component (modeling continuous fluctuations), and a jump component (capturing sudden events such as breakdowns or recoveries in trust).

Together, these two representations are complementary and jointly enable both rigorous theoretical analysis and accurate empirical simulation of trust dynamics in strategic interactions.

*A Lévy Process Approach to Modeling Trust Dynamics*

In order to capture the dynamic evolution of trust between agents over time, we model the trust level $X_t$ at time t as a Lévy process:

$$X_t = X_0 + \gamma t + \sigma W_t + \int_0^t \int_\mathbb{R} z \, \tilde{N}(ds, dz) \tag{5}$$

where:

- $X_0$ denotes the initial level of trust.
- $\gamma \in \mathbb{R}$ is the drift rate, representing the overall directional tendency of trust changes (e.g., the system's inclination to increase or decrease trust in the absence of noise).
- $\sigma \in \mathbb{R}$ quantifies the intensity of continuous Gaussian noise, modeling typical fluctuations in trust.
- $W_t$ is a Wiener process (Brownian motion) representing continuous white noise.
- $\tilde{N}(ds, dz)$ is the compensated jump counting measure constructed from the Lévy measure, defined as

$$\tilde{N}(ds,dz) = N(ds,dz) - v(dz)ds \tag{6}$$

where:

- $N(ds,dz)$ is a Poisson random measure on time and jump size space, and $v(dz)$ is the Lévy measure characterizing the intensity and distribution of jump magnitudes.

In this model, the first term $(\gamma t)$ captures the linear trend and the overall direction of trust over time. The second term $(\sigma W_t)$ accounts for small, continuous day to day fluctuations, such as ordinary uncertainty or the influence of minor events. The third component the double integral models discontinuous, abrupt jumps that represent sudden shocks, such as political scandals, betrayal, or other impactful events that markedly increase or decrease trust.

As discussed earlier, each term in Equation (5) represents a specific type of dynamic behavior in the evolution of trust over time. In particular, the double integral at the end of the expression, namely:

$$\int_0^t \int_\mathbb{R} z\, \tilde{N}(ds,sz) \tag{7}$$

carries a particularly critical meaning, which will be elaborated on in detail in the following discussion.

*Lévy Measure*

The Lévy measure, denoted by $v(dz)$, plays a central role in both the definition of the characteristic function and the structure of the jump component in the model. It provides a rigorous mathematical tool for quantifying the frequency and magnitude of discontinuous jumps in the stochastic process. Essentially, the Lévy measure specifies how often random shocks of varying intensities ranging from minor to extreme occur. Within the context of trust dynamics, it captures the statistical signature of rare but consequential events, such as the exposure of a betrayal or an abrupt reversal in the counterpart's behavior, which can lead to sharp changes in trust.

The symbol $\tilde{N}(ds, dz)$ denotes the compensated Poisson random measure, which models the number of jumps of size $z$ occurring at time $s$. This measure is adjusted such that its expected value is zero hence the term compensated. This adjustment renders the jump process statistically homogeneous and dynamically stable over time, thereby enabling more precise analytical treatment within the model.

The jump integral represents the cumulative impact of all discrete jumps occurring within the time interval $[0, t]$. In our model, this term captures the total contribution of sudden, discontinuous changes in trust over time. Such jumps may reflect psychological shocks, revelations of hidden information, unexpected decisions, or significant events within the interaction between players. By aggregating these discontinuities, the integral provides a rigorous mathematical account of rare but influential events that shape the trust dynamics.

Practically, this integral enables the simulation of dynamics that combine both gradual trends and abrupt jumps. In behavioral models such as the trust game, this duality is essential for accurately capturing human reality: individuals may experience a gradual erosion of trust or, alternatively, a sudden shift triggered by a specific shock such as the revelation of deceit that instantaneously alters their trust level.

***Social Interpretation of the Model***

Within this framework:

- A large value of $\sigma$ indicates that the society is highly sensitive to small fluctuations, reflecting unstable trust dynamics.
- If the Lévy measure $\nu(dz)$ places significant mass on large values of $z$, it implies that the society is prone to sudden and severe shocks in trust.
- A positive drift parameter $\gamma$ suggests a natural tendency for trust to increase over time, whereas a negative value reflects a gradual erosion of trust.

*Key Advantages of the Lévy Process Model*

In analyzing the dynamics of trust, the Lévy process employed in this model offers fundamental advantages over classical approaches. Below, we outline three main features of the model along with detailed explanations, highlighting the clear distinctions from traditional methods:

*Nonlinear and Non-Normal Behavior*

In many classical models, such as those based on Brownian motion (Wiener process), it is assumed that changes in trust occur continuously, gradually, and follow a normal distribution. However, human behavior toward trust often deviates from these assumptions. Individuals may abruptly and dramatically lose trust in response to sudden shocks or experience intense behavioral fluctuations that are incompatible with a normal distribution.

The Lévy process model effectively addresses this issue through its jump components, derived from a stochastic Poisson process. Unlike normal models, the Poisson process allows for discrete and rare changes, which are infrequent but can have a significant impact. Put simply, the Lévy model can incorporate "rare events with large consequences" phenomena that play a crucial role in human relationships and trust dynamics (e.g., the revelation of a major lie or an emotional shock).

*Trust Recovery via the Gaussian Component*

The Wiener component of the model, represented by $\sigma W_t$, captures small and continuous noise fluctuations. These fluctuations can reflect the cumulative effects of everyday positive interactions, minor acts of goodwill, and incremental efforts to rebuild trust. The inclusion of this component endows the model with the capacity for gradual trust recovery an aspect commonly observed in real human relationships.

Importantly, in the absence of sudden jumps, the Lévy process reduces to Brownian motion, thereby exhibiting smooth and predictable behavior under calm conditions.

*Integration of Continuous and Discontinuous Components: A More Realistic Representation*

A key advantage of the Lévy model lies in its unified framework that encapsulates three distinct types of dynamics:

- A deterministic drift component ($\gamma t$) representing the overall trend of trust growth or decline;
- A continuous stochastic component (Gaussian noise) accounting for everyday fluctuations;
- A discontinuous jump component modeling sudden and significant shocks.

This integration enables the model to capture both smooth, gradual behaviors and abrupt, severe changes, thereby providing a more realistic and adaptable representation of human and social trust dynamics.

To illustrate the core differences between continuous and jump driven stochastic behaviors, we simulate and present representative sample paths from both Brownian motion (Wiener process) and a Lévy process. These trajectories help clarify the modeling power of each stochastic framework in capturing distinct types of behavioral dynamics.

As shown in Fig 1, the Brownian motion generates smooth and continuous sample paths, characterized by gradual variations over time and the absence of abrupt changes. Such processes are widely employed for modeling behavior under Gaussian noise where fluctuations evolve continuously and no large external shocks occur. This makes them particularly suitable for phenomena involving steady decision-making or environments without abrupt disruptions.

In contrast, Fig 2 displays ten sample trajectories from a Lévy process that integrates both continuous Brownian fluctuations and sudden, randomly occurring jumps, modeled via a Poisson process with normally distributed jump sizes. These jumps create sharp discontinuities, enabling the process to capture abrupt behavioral shifts, such as emotional surges or sudden collapses in trust. Unlike Brownian motion, the Lévy framework allows for a more faithful modeling of real world decision dynamics under uncertainty, particularly when sudden shocks play a central role.

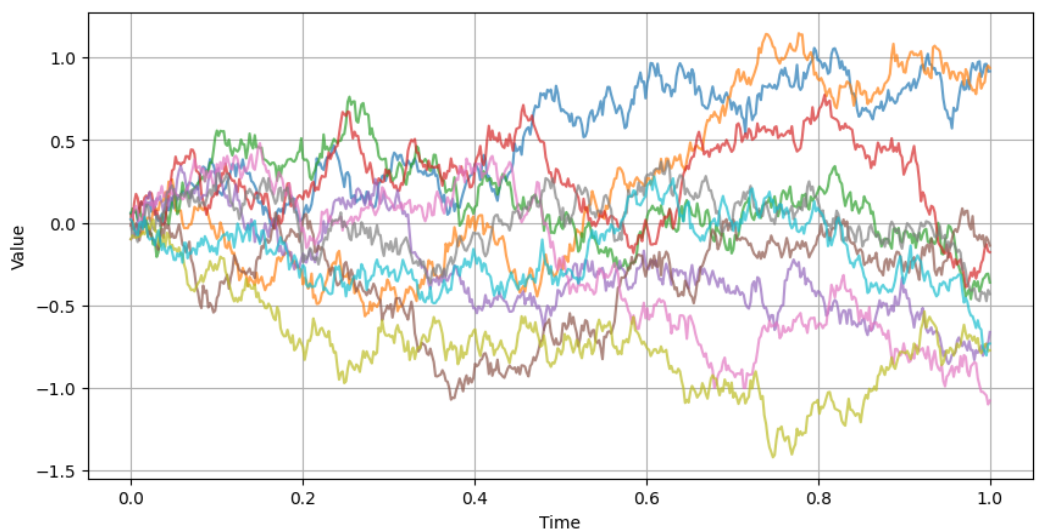

Figure 1.Ten simulated sample paths of a Brownian motion. The trajectories exhibit continuous and smooth variations, illustrating the behavior of a purely Gaussian process without sudden jumps.

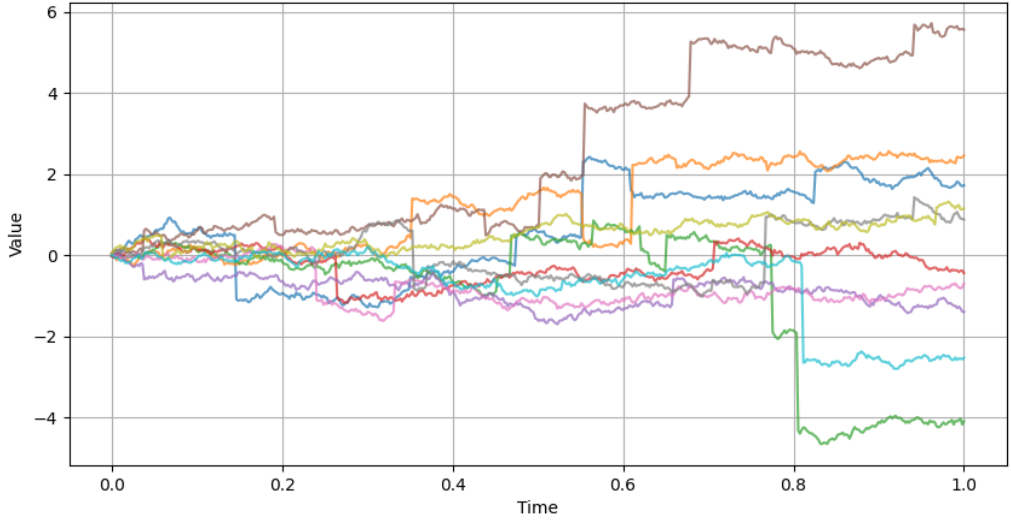

Figure 2.Ten sample paths of a Lévy process with Gaussian jumps. The presence of discontinuous spikes reflects abrupt behavioral shifts induced by stochastic jumps.

## *Modeling Trust Game Dynamics via Lévy Processes: Parameter Roles and Simulation*

We aim to model the trust level as a stochastic process $X_t$, which represents the trust level of the first player in the Trust Game at time $t$. To this end, we employ a Lévy process characterized by the following general form:

$$X_t = X_0 + \gamma t + \sigma W_t + \sum_{i=1}^{N_t} J_i \tag{8}$$

Precise definitions of each component are as follows:

| Symbol | Meaning | Description |
| --- | --- | --- |
| $X_t$ | Trust level at time $t$ | State variable representing the system's trust level at time $t$ |
| $X_0$ | Initial trust value | Starting point of the process |
| $\gamma$ | Drift rate | Long-term directional trend of the trust process; positive values indicate increasing trust |
| $\sigma$ | Gaussian noise intensity | Magnitude of small random fluctuations modeled via Brownian motion |

| $W_t$ | Standard Brownian motion | Models continuous small and incremental changes in trust (white noise) |
|---|---|---|
| $N_t$ | Number of jumps up to time $t$ | A Poisson process with rate $\lambda$, governing the frequency of jumps |
| $J_t$ | Size of the $i$-th jump | Random variable representing sudden changes in trust (e.g., a major betrayal or sudden reward) |

Table 1. Description of model parameters in the Lévy-based trust process. Each symbol represents a component of the stochastic process $X_t$, modeling the evolution of trust over time in the Trust Game.

### *The Role of Jumps and the Poisson Process in Modeling Trust Dynamics*

In the proposed model, the term

$$\sum_{i=1}^{N_t} J_i \qquad (9)$$

represents the cumulative effect of sudden shocks (or stochastic jumps) in the level of trust up to time t. These jumps introduce discontinuous changes in trust, potentially arising from unexpected events, irrational decisions, or structural disruptions in the interaction between the two players.

The number of such jumps occurring by time t is captured by the random variable $N_t$, which follows a homogeneous Poisson process with intensity $\lambda$:

$$N_t \sim Poisson(\lambda t) \qquad (10)$$

In other words, $N_t$ denotes the number of jumps that have occurred up to time t, and each jump is represented by a random variable $J_i$ for $i = 1, 2, \ldots, N_t$. Thus, the term (10) constitutes a random sum in which both the number of terms and the magnitude of each term are stochastic. This structure renders the jump dynamics inherently random and fundamentally unpredictable.

The homogeneous Poisson process inherently assumes that the inter-arrival times between successive jumps are independent and exponentially distributed with parameter

$\lambda$. This parameter $\lambda$ represents the average rate at which jumps occur per unit of time. In other words:

- A larger $\lambda$ implies a higher expected number of jumps within a given time interval.
- An increase in $\lambda$ corresponds to more frequent and intense fluctuations in trust.
- This parameter governs the intensity of sudden events or trust shocks and plays a critical role in shaping the dynamics of the model.

In modeling trust dynamics using jump processes, each sudden jump occurring at time $t$ t must not only have a well defined occurrence time captured by the counting process $N_t$ but also a magnitude or intensity that is of critical importance. To represent this, we employ the random variables $J_i$.

Each $J_i$ denotes the magnitude of the sudden change in trust associated with the i-th jump. In other words, if a shock or event occurs in the system at a given moment, $J_i$ determines the impact of this shock on the trust level, which can be either positive (an increase in trust) or negative (a decrease in trust).

These variables are typically assumed to be independent and identically distributed (i.i.d.) and follow a specified distribution, such as a normal distribution with mean $\mu_J$ and variance $\sigma_J^2$:

$$J_i^i \sim \qquad ) \tag{11}$$

The statistical properties of $J_i$ (such as its mean and variance) can be determined based on empirical evidence or the expected characteristics of trust dynamics:

- A positive mean $\mu_J$ indicates that, on average, jumps tend to increase trust.

- A negative mean suggests that the system experiences sudden collapses in trust on average.

- A high variance $\sigma_J^2$ reflects considerable uncertainty in the magnitude of shocks.

This structure enables the model to simultaneously capture both smooth, gradual fluctuations (via the Wiener process) and sudden, discrete jumps (via the $J_i$ terms) in trust dynamics.

In the context of the Trust Game, we define shocks as: "Sudden, unpredictable, and discrete changes in the behavior of Player 2, which can exert either positive or negative effects on the trust level of Player 1."

In typical interactions, Player 2 may follow a relatively stable behavioral pattern. However, in real-world scenarios, unexpected deviations frequently occur:

- At times, Player 2 may suddenly exhibit a high level of generosity (positive shock).
- At other times, they may abruptly become untrustworthy, greedy, or unreliable (negative shock).

Each shock formally represented as a jump in the process corresponds to an instantaneous change in the level of trust.

- If the shock magnitude is positive, trust increases sharply.
- If negative, trust declines suddenly.

In our model:

- $\lambda$ denotes the average frequency of such shocks per unit time.

- $\mu_J$ reflects the directional bias of these shocks whether they are predominantly positive or negative.
- $\sigma_J$ captures the dispersion of their magnitudes indicating whether the behavioral deviations are mild or extreme.

Thus, in this Lévy based framework, shocks represent abrupt behavioral changes in Player 2 that occur outside the regular trajectory of trust dynamics. These events may vary in scale, sign, and predictability, and can significantly influence the decision-making of Player 1 in either constructive or detrimental ways.

**Results**

To simulate the trust game dynamics using the Lévy process model, we consider four distinct scenarios:

(1) Increasing trust driven by positive shocks
(2) Unstable Trust with Scattered Shocks
(3) Declining trust accompanied by high volatility
(4) Trust dynamics influenced by unpredictable behavior of the counterpart

*Increasing trust driven by positive shocks*

To simulate this scenario, we specify a set of parameters for the Lévy driven trust process. The parameters are selected to reflect a context in which trust is gradually increasing due to infrequent but significant positive shocks. The following table provides the numerical values used in the simulation, their mathematical role in the Lévy process, their function in the trust game setting, and a behavioral interpretation of each parameter.

| Parameter | Value | Role in the Mathematical Model | Role in the Trust Game | Behavioral and Real-World Interpretation |
|---|---|---|---|---|
| $X_0 = 0.5$ | Initial value | Initial condition of trust process $X_t$ | The game starts with a moderate level of trust | Player 1 transfers half of their endowment to Player 2 |
| $\gamma = 0.03$ | Positive | Drift – average rate of change | Intrinsic tendency for trust to build over time | On average, Player 2 behaves well, gradually fostering trust |
| $\sigma = 0.1$ | Moderate | Volatility – Gaussian noise level | Behavioral fluctuations or uncertainty in interpreting actions | Trust fluctuates but generally follows the direction set by drift and shocks |
| $\lambda = 0.3$ | Low | Rate of Poisson jumps | Rare but significant behavioral events by Player 2 | Occasionally, Player 2 behaves exceptionally well, causing sudden increases in trust |
| $\mu_J = 0.2$ | Positive | Mean size of jump (positive direction) | Jumps tend to favor trust | Player 2 exhibits unexpectedly generous behavior at times |
| $\sigma_J = 0.05$ | Small | Standard deviation of jump sizes | Jumps are similar in magnitude | Sudden behavioral shifts are usually of comparable and predictable magnitude (e.g., consistent generosity) |

Table 2. Description of model parameters in the Lévy-based trust process. Each symbol represents a component of the stochastic process, modeling the evolution of trust over time in the Trust Game.

To illustrate the dynamics of the increasing-trust scenario, we numerically simulate the Lévy-driven trust process using the parameter set specified in Table 2. The simulation captures the temporal trajectory of trust under conditions of mild environmental volatility, infrequent but meaningful positive jumps, and a consistently optimistic behavioral drift. The resulting path reflects how stable trust may emerge and grow in response to predominantly cooperative behavior from the counterpart. Figure 3 presents the time series of trust level $X_t$ generated by this simulation.

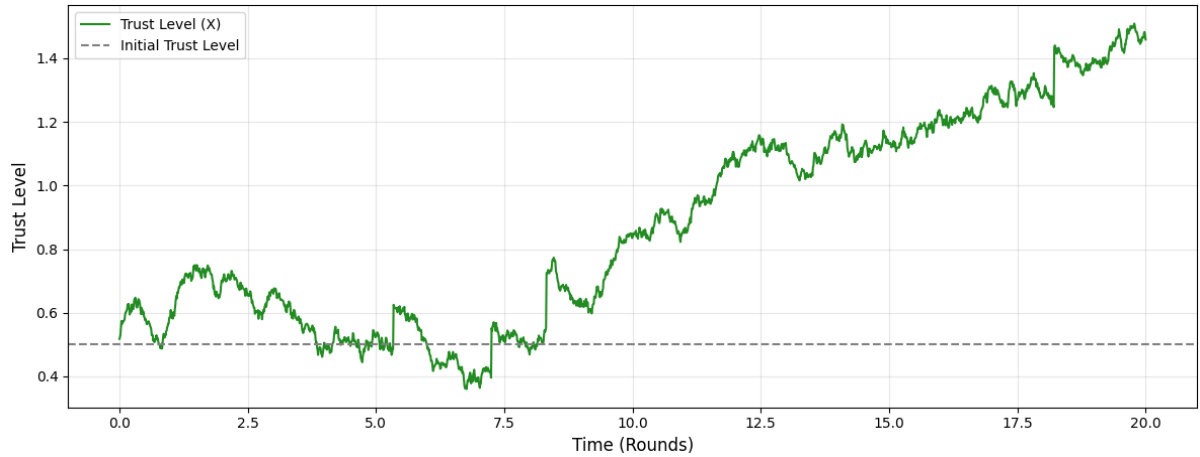

Figure 3. Simulated trajectory of trust level $X_t$ under the Lévy process, illustrating gradual trust formation driven by positive drift and infrequent upward jumps.

To illustrate the dynamics of increasing trust under a Lévy process, we conducted a numerical simulation based on the parameter values summarized in Table 2. The initial trust level was set at a moderate value $X_0 = 0.5$, with a positive drift $\gamma = 0.03$, reflecting a general tendency toward growing trust. Random jumps, modeled with a low intensity $\lambda = 0.3$ and a positive mean size $\mu_J = 0.2$, represent occasional acts of exceptional generosity or unexpectedly favorable responses by Player 2. Environmental noise with moderate variance $\sigma = 0.1$ accounts for ongoing fluctuations. This scenario captures the gradual and stable formation of trust in a dyadic interaction where Player 2 exhibits predominantly positive behavior, occasionally reinforced by rare but impactful actions.

*Behavioral Dynamics of Players in the Lévy-Process-Based Trust Game under the Scenario of Trust Enhancement Driven by Positive Shocks*

Player 1 (The Trustor): At the beginning of each round, Player 1 decides the amount of capital to send. In this scenario, since the initial trust level is set at 0.5 (i.e., half of the initial endowment), Player 1 exhibits a relatively trustful behavior.

Player 2 (The Trustee): Upon receiving the funds, Player 2 decides how much to return. Given that $\gamma > 0$, Player 2 tends to behave in a trustworthy manner. Moreover, due to $\mu_J > 0$ and $\lambda > 0$, Player 2 occasionally exhibits exceptionally positive behaviors (e.g., returning more than expected), resulting in upward jumps in the trust level.

Over time: Trust increases because:

(1) $\gamma > 0$: The average behavior of Player 2 is reliable.

(2) $\lambda > 0$ and $\mu_J > 0$: Positive shocks intermittently cause sudden jumps in trust.

(3) $\sigma$: Introduces fluctuations, but its long-term effect is dominated by drift and jump components.

The trajectory of $X_t$ gradually grows with upward jumps, reflecting the stable formation of trust between the two players throughout repeated interactions.

*Unstable Trust with Scattered Shocks*

To examine how trust evolves in environments with erratic and unpredictable behavior, we simulate the trust game under a parameter setting that reflects high environmental volatility and frequent but neutral-impact shocks. The table below outlines the selected parameters and their interpretations in both the mathematical model and the behavioral dynamics of trust.

| Parameter | Value | Role in the Mathematical Model | Role in the Trust Game | Behavioral and Real-World Interpretation |
|---|---|---|---|---|
| $X_0 = 0.5$ | Initial | Starting level of trust | Player 1 sends half of the initial endowment | Trust begins at a moderate level |
| $\gamma = 0$ | Zero | No deterministic trend in trust | Player 2 is neither consistently trustworthy nor untrustworthy | No long-term tendency to increase or decrease trust |
| $\sigma = 0.2$ | High | Significant Gaussian noise | Substantial randomness in interactions | High uncertainty in the environment or social context |
| $\lambda = 0.7$ | High | High jump frequency | Frequent behavioral surprises from Player 2 | Unpredictable behavior occurs often |
| $\mu_J = 0$ | Neutral | Mean of jumps is zero | Jumps do not have a systematic positive or negative effect | On average, shocks are neither good nor bad |
| $\sigma_J = 0.15$ | High | High variability in jump sizes | Behavior of Player 2 ranges from highly positive to highly negative | Trust can suddenly rise or fall due to extreme but neutral shocks |

Table 3.Parameter configuration for simulating the scenario of "Unstable Trust under Dispersed Shocks" This table presents the parameter setup employed to simulate the dynamics of trust in an environment characterized by high environmental volatility and frequent, behaviorally neutral shocks.

To explore the dynamics of Unstable Trust with Scattered Shocks, we conducted numerical simulations using the parameter values outlined in Table 3. The resulting trajectory of $X_t$ the trust level is presented in Figure 4.

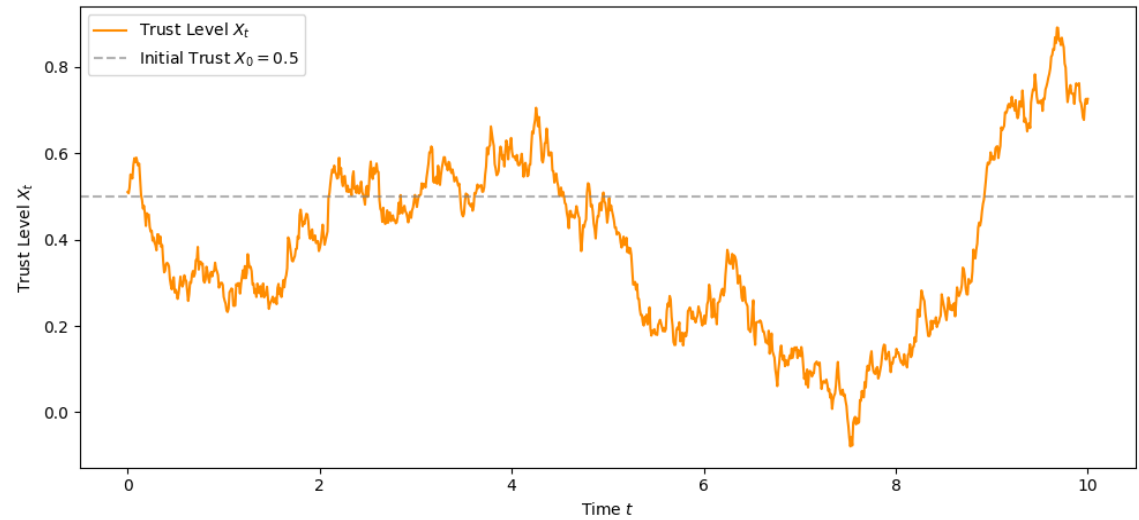

Figure 4.Temporal evolution of trust simulated under the parameter configuration of Scenario 2, titled "Unstable Trust with Scattered Shocks," as detailed in Table 3.

Figure 4 illustrates the temporal evolution of the trust level $X_t$ within the Lévy process-based model under the "Unstable Trust with Scattered Shocks" scenario. In this setting, trust starts at a moderate initial value $X_0 = 0.5$ with no intrinsic directional drift ( $\gamma = 0$ ). Instead, significant environmental volatility ( $\sigma = 0.2$ ) combined with frequent yet directionless shocks ( $\lambda = 0.7$, $\mu_J = 0$, $\sigma_J = 0.15$ ) induces unpredictable fluctuations in trust. This scenario captures interactions where the erratic and unstable behavior of Player 2 hinders the formation of stable trust over time.

*Behavioral Dynamics of Players in the Lévy-Process-Based Trust Game under the Scenario of Unstable Trust with Scattered Shocks*

Player 1 (Trustor): Initiates the interaction with a moderate trust level of $X_0 = 0.5$, yet exhibits no inherent tendency toward either increasing or decreasing trust, as indicated by a zero drift parameter ( $\gamma = 0$ ).

Player 2 (Trustee): Displays highly erratic and volatile behaviour sometimes cooperative, sometimes uncooperative, without any discernible pattern. The high frequency of shocks ( $\lambda = 0.7$ ) combined with substantial variability in their magnitude ( $\sigma_J = 0.15$ ) introduces significant fluctuations and instability into the trust dynamics.

The result is that the trust level exhibits unpredictable oscillations over time, without converging toward any stable or increasing trajectory. This outcome reflects an interaction in which trust fails to consolidate due to the counterpart's erratic and inconsistent behavior.

*Declining trust accompanied by high volatility*

To analyze the dynamics of declining trust under conditions of pronounced volatility, we simulate the trust game using a parameter configuration representative of this scenario. The table below presents the key parameters alongside their roles within the mathematical framework and their behavioral and real-world interpretations.

| Parameter | Value | Role in the Mathematical Model | Role in the Trust Game | Behavioral and Real-World Interpretation |
|---|---|---|---|---|
| $X_0 = 0.7$ | High | Initial trust level | Player 1 starts with relatively high trust | The trustor commits a substantial portion of capital at the outset |
| $\gamma = -0.02$ | Negative | Intrinsic tendency toward decreasing trust | Player 2 gradually exhibits untrustworthy behavior | The trustee's behavior increasingly undermines trust over time |
| $\sigma = 0.2$ | High | Significant environmental volatility | External conditions and interactions are highly unstable | The environment and social context are markedly turbulent and unstable |
| $\lambda = 0.8$ | Relatively high | High frequency of shocks | Unexpected behaviors by Player 2 occur frequently | The trustee's unpredictable actions are common and recurrent |
| $\mu_J = -0.2$ | Strongly negative | Mean of shocks strongly decreasing | Shocks predominantly cause further declines in trust | Negative shocks regularly reduce the trust level significantly |
| $\sigma_J = 0.1$ | High | High dispersion in shock magnitudes | Player 2's behavior varies from highly detrimental to mildly negative | The trustee's actions fluctuate between severely harmful and slightly adverse |

Table 4. Summary of key parameters and their respective roles in modeling declining trust under high volatility conditions.

Based on the parameter configuration summarized in Table 4, we conducted a simulation of the trust dynamics within the declining trust scenario characterized by high volatility. The following figure presents the resulting temporal trajectory of trust, illustrating the complex fluctuations and instability inherent in this setting.

In this scenario, Player 1 enters the game with a relatively high initial trust level ($X_0 = 0.7$), indicating a willingness to allocate a substantial portion of their initial capital to Player 2. However, over the course of 20 interaction rounds, the overall trajectory of trust exhibits a declining trend.

This decline arises from the interplay of three key factors:

- Negative drift ($\gamma = -0.02$): This parameter represents Player 1's intrinsic tendency to reduce trust, typically reflecting their perception of opportunistic or untrustworthy behavior by Player 2 over time.

- High environmental volatility ($\sigma = 0.2$): These fluctuations signify instability within the interaction environment or ambiguous signals received from Player 2. This factor induces continual oscillations and instability in the trust level.
- Frequent and severe negative shocks ($\lambda = 0.8$, $\mu_J = -0.2$): These parameters model sudden, impactful behaviors by Player 2 that significantly undermine trust. The high variance in shock magnitude ($\sigma_J = 0.1$) indicates that these negative behaviors range from minor setbacks to major betrayals.

Together, these elements generate an unstable, fluctuating, yet overall declining trust trajectory. Although intermittent environmental volatility occasionally leads to temporary recoveries in trust, the trust level never returns to its initial value. This pattern reflects situations where Player 1, upon repeatedly observing signs of non-commitment from Player 2, gradually loses trust and ultimately disengages from continued cooperation.

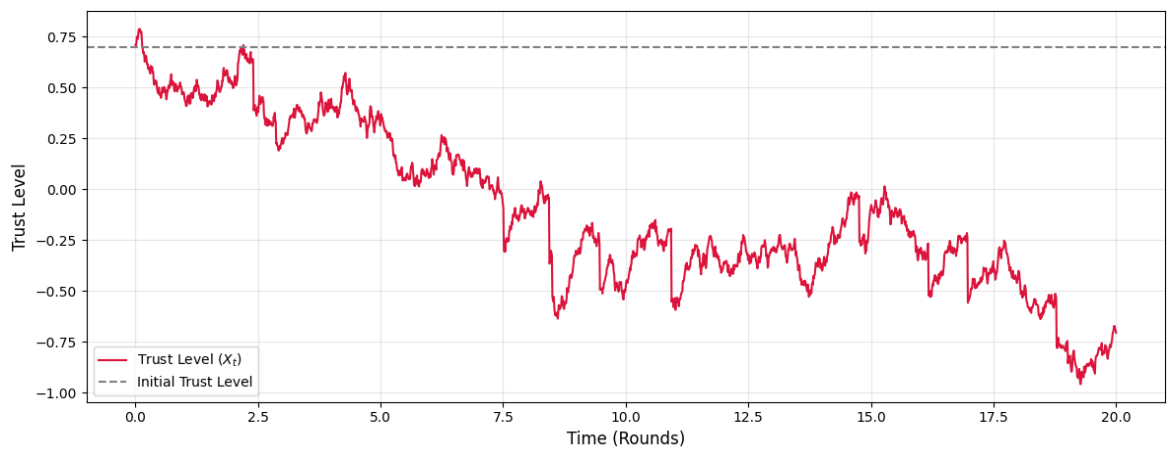

Figure 5. Simulated temporal trajectory of trust depicting pronounced fluctuations and an overall downward trend under the declining trust scenario with elevated volatility.

*Trust dynamics influenced by unpredictable behavior of the counterpart*

In real-world social interactions, one of the most challenging situations arises when:

- The counterpart exhibits no stable behavioral pattern;
- Actions alternate between highly generous and outright betrayals;

- Trust neither steadily increases nor decreases, but instead fluctuates in an unpredictable manner.

Such a situation is well-captured by Lévy processes with zero drift, high volatility, and variable jumps, allowing us to simulate erratic and unstable trust dynamics. To model this scenario within the Lévy-based framework of the trust game, we define the following parameters in Table 5.

| Parameter | Numerical Value | Role in the Trust Game | Behavioral and Real-World Interpretation |
|---|---|---|---|
| $X_0 = 0.5$ | Moderate | Initial level of trust | Player 1 starts with moderate trust and invests half of the endowment with caution. |
| $\gamma = 0$ | Zero | No intrinsic drift in trust | Player 2 has no consistent behavioral tendency—neither trustworthy nor untrustworthy on average. |
| $\sigma = 0.3$ | Very high | Strong environmental volatility | The interaction environment is tense and unstable; decisions are highly susceptible to unpredictable factors. |
| $\lambda = 1.0$ | Very high | High frequency of shocks | Player 2 frequently exhibits sudden, erratic behaviors. |
| $\mu_J = 0.0$ | Zero | Neutral average jump size | Shocks are neutral in the long run, but may appear extremely positive or negative in the short term. |
| $\sigma_J = 0.2$ | Very high | High variability in shock magnitude | Player 2 may behave extremely generously or maliciously without following any discernible pattern. |

Table 5. Parameter configuration of the Lévy-based model for simulating erratic and unstable behavioral dynamics. This setup captures a scenario in which trust is subject to intense fluctuations and alternating behavioral shocks, with no clear upward or downward trend—mirroring real-world conditions of unpredictability and instability in interpersonal interactions.

Building upon the parameter configuration detailed in Table 5, we conducted a simulation to explore how trust evolves under conditions of extreme behavioral unpredictability. This scenario, characterized by the absence of stable behavioral patterns and sharp fluctuations in actions, allows us to examine the dynamic consequences of instability and irregularity in trust formation. The figure below illustrates the simulated trajectory of trust over time.

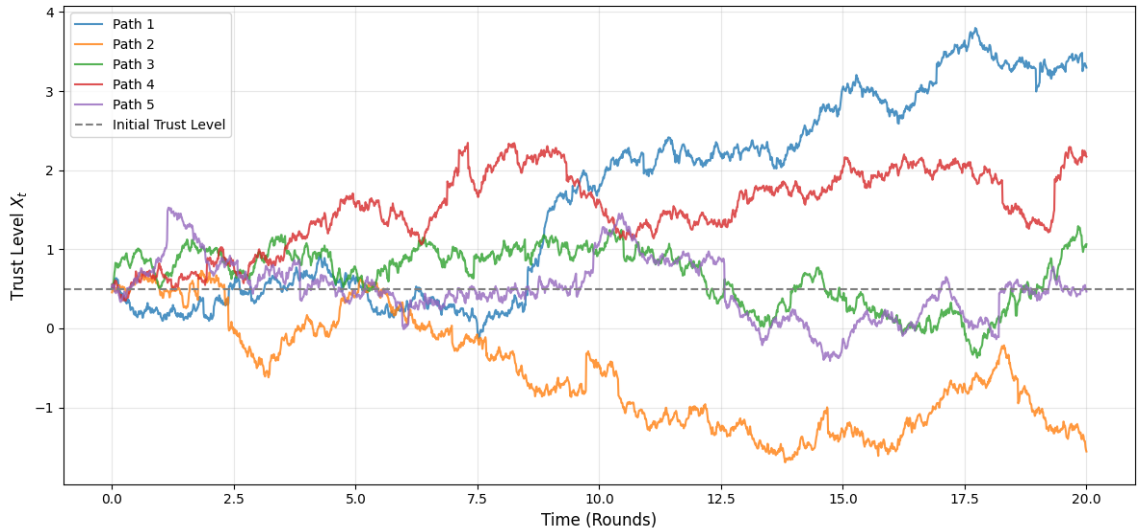

Figure 6.Simulated Temporal evolution of trust level $X_t$ in the "Unpredictable Behavior of Player 2" scenario.Simulated trust trajectory under highly erratic behavioral conditions, illustrating sharp oscillations and the absence of a discernible long-term trend.

As illustrated in the simulated figure, the panel displays five independent realizations of the trust process, all initiated from the same initial condition $X_0 = 0.5$. The model parameters include elevated environmental volatility ($\sigma = 0.3$), a high frequency of behavioral shocks ($\lambda = 1.0$), and considerable dispersion in shock magnitudes ($\sigma_J = 0.2$), with the mean shock size set to zero ($\mu_J = 0.0$).

These conditions give rise to heterogeneous and sometimes divergent trust trajectories: some paths exhibit trust growth despite volatility; others collapse under repeated negative experiences; and several oscillate near the initial level without a clear trend. This simulation captures the dynamics of environments where the counterpart's behavior lacks stable patterns, resulting in pronounced instability in trust development.

*Analysis of the Trust Trajectories Presented in Figure 6*

Path 1 (Blue): This trajectory exhibits an initially mild fluctuation followed by a sharp and sustained increase in trust beginning around t = 8. In real-world terms, this may represent interactions in which Player 2 despite an overall erratic behavioral profile randomly exhibits trustworthy behavior. The repetition of such behavior reinforces Player 1's trust, ultimately leading to a significant upward shift.

Path 2 (Orange): A continuous decline in trust characterizes this critical scenario, where the trust level follows a downward trend from the beginning and drops below -1 by the end of the game. This reflects situations in which Player 2 repeatedly exhibits negative random behaviors, resulting in a complete erosion of trust.

Path 3 (Green): This path displays moderate fluctuations, with a gradual initial decline that remains relatively stable. However, in the second half of the interaction, random behavioral shocks lead to a slow erosion of trust. It typifies interactions where the counterpart displays no discernible pattern alternating between constructive and destructive behaviors without fostering either robust trust growth or total breakdown.

Path 4 (Red): Characterized by a gradual rise with intermittent positive jumps, this path shows a slow but consistent increase in trust, punctuated by several upward shocks. It demonstrates that even within an unstable environment, trust can grow if positive shocks occur at critical junctures even when the system's baseline tendency is neutral.

Path 5 (Purple): This trajectory oscillates persistently around the initial trust level, mostly hovering near $X_0 = 0.5$. It corresponds to cases where Player 2's behavior is entirely patternless, producing a balanced mix of positive and negative actions.

Even under the assumption of zero-mean shocks, the trust process can diverge significantly either growing sharply or collapsing due to the combination of high environmental volatility ($\sigma$) and dispersed jump magnitudes ($\sigma_J$). This mechanism highlights how behavioral instability in Player 2 can give rise to highly divergent trust outcomes.

Overall, the model illustrates that in socially unstable and behaviorally unpredictable environments, trust can either flourish or disintegrate purely by chance even when the system appears neutral at first glance.

*Structural Instability in Trust under Stochastic Behavior of the Second Player: A Lévy Path Modeling Approach*

In Figure 7, the trust trajectories are simulated 100 times under the scenario of unpredictable behavior by the second player. The model parameters include high volatility ($\sigma = 0.3$), a high rate of jump occurrence ($\lambda = 1.0$), and a zero mean jump magnitude ($\mu_J = 0.0$). The colored lines represent individual realizations of the stochastic process, while the black line shows the average of all trajectories. The surrounding grey band indicates one standard deviation ($\pm\sigma$), capturing the degree of behavioral dispersion.

Although the average trajectory remains oscillatory and close to the initial trust level, some paths exhibit sharp declines, while others experience unexpected increases in trust. This illustrates that the unpredictable behavior of the second player does not lead to a stable pattern and, depending on random events, can generate highly divergent trust outcomes. This modeling captures the essence of human interactions in environments characterized by uncertainty.

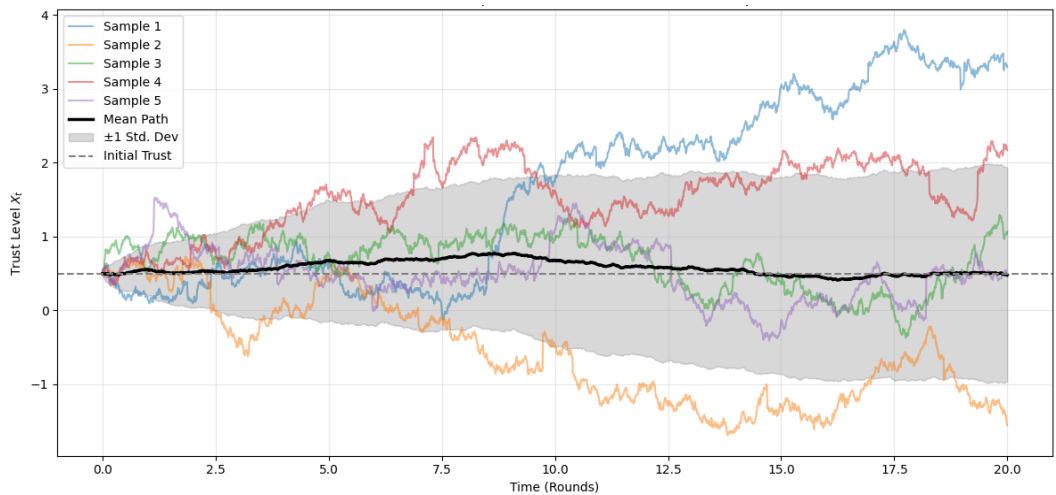

Figure 7. Simulation of trust levels under unpredictable behavior of the second player; displaying the mean trajectory, sample paths, and volatility band based on the Lévy process model.

As shown in Figure 7, although the average of the trust trajectories remains close to the initial level without exhibiting a clearly increasing or decreasing trend the high dispersion among individual paths indicates that the second player's behavior induces structural instability in trust. In other words, even when the shocks are on average neutral, the combination of high volatility and frequent occurrence of unexpected events generates diverging behavioral patterns: some paths exhibit sharp declines in trust, others display temporary surges, and some fluctuate around a moderate level. This simulation illustrates that, in the absence of trust reinforcing mechanisms, random and erratic behaviors can keep trust in an unstable state, even if the "average trust" appears unchanged.

*Three-dimensional analysis of the impact of shock mean and occurrence rate on the final level of trust in the Lévy based model*

Given that our model is built upon a Lévy process characterized by two fundamental components: continuous fluctuations (Brownian motion) and sudden jumps (random shocks) each of these components captures distinct behavioral and environmental factors in the dynamics of the trust game. While classical models of trust often fail to incorporate such dynamics, our approach enables the simulation of unpredictable and structurally unstable behaviors of the second player. To better understand the explanatory power of the proposed model, we analyze how variations in key parameters such as volatility, shock frequency, and average shock magnitude affect the final level of trust.

In modeling the trust game through a Lévy process, one of the most crucial components is the second player's unexpected behavior, represented by stochastic shocks drawn from a normal distribution. These shocks are parametrized by their rate of occurrence and average intensity. Accordingly, to explore how these shocks contribute to the formation or erosion of trust, the three-dimensional diagram below illustrates the joint effect of shock frequency ($\lambda$) and average shock magnitude ($\mu_J$) on the final trust level after 20 rounds of interaction.

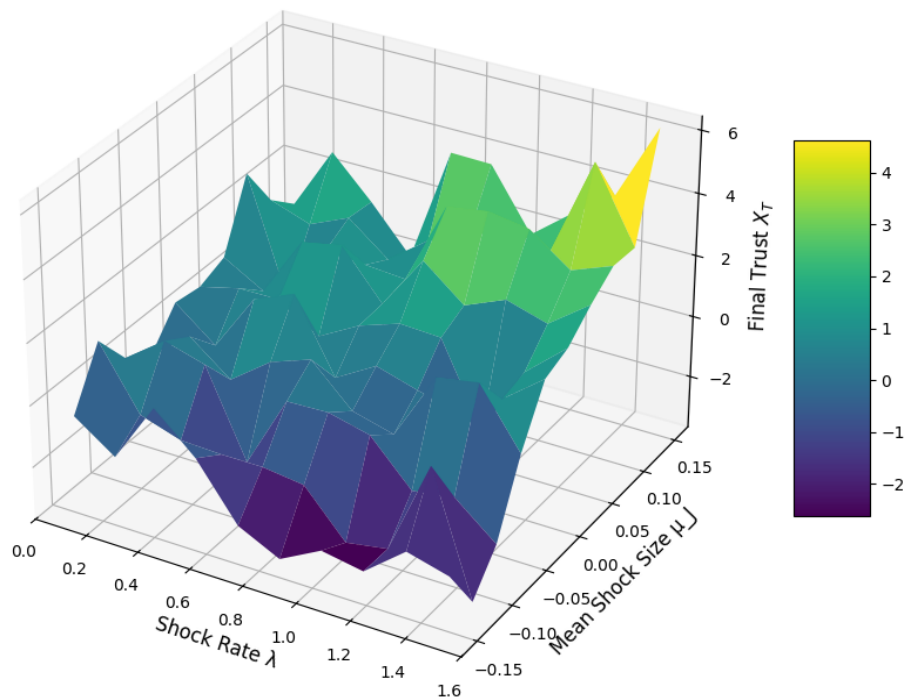

Figure 8. Variation in the final trust level ($X_t$) as a function of the average shock magnitude ($\mu_J$) and their occurrence rate ($\lambda$) in the Lévy-based model.

As previously noted, in modeling trust using the Lévy process, shocks represent the sudden, unexpected, and random behaviors of the second player, which play a pivotal role in either building or undermining trust. Two key parameters in this context are:

- $\mu_J$ – Mean of Shocks:

    This parameter reflects the overall tendency of the second player in their sudden behavioral deviations. A negative value indicates predominantly untrustworthy or trust-reducing behaviors, while a positive value reflects unexpected generosity or actions that abruptly enhance trust.

- $\lambda$ – Shock Frequency (Rate of Occurrence):

    This parameter determines how often the second player exhibits unexpected behavior. Higher values of $\lambda$ imply more frequent behavioral disruptions and greater instability, whereas lower values suggest more stable, consistent, and predictable conduct by the second player.

Figure 8 illustrates that when the mean of shocks ($\mu_J$) is positive indicating that the second player's sudden behaviors are, on average, trust-enhancing the final level of trust tends to be higher. This effect is especially pronounced when the shock frequency ($\lambda$) is also high, as a greater number of positive behaviors accumulate over time, reinforcing trust. In contrast, when $\mu_J$ is negative, the shocks have a detrimental impact. Even at low frequencies (i.e., small $\lambda$), the final trust level may significantly decline. In such cases, trust becomes fragile, since even infrequent negative shocks can have lasting adverse effects.

When $\mu_J$ is close to zero (i.e., shocks are on average neutral), the level of trust becomes highly sensitive to the value of $\lambda$. A high shock frequency in this scenario leads to pronounced fluctuations and instability in the trust trajectory, potentially resulting in either an increase or decrease in trust, depending on the random sequence of shocks.

The simulation results in Figure 8 suggest that the interplay between the frequency of shocks ($\lambda$) and their directional tendency ($\mu_J$) plays a decisive role in the emergence, persistence, or erosion of trust. This analysis underscores that the mean of shocks alone is not sufficient for understanding trust dynamics; the rate at which such shocks occur must also be taken into account.

To further investigate the dynamics of trust, we simulated the variations in trust level $X_t$ as a function of both the drift rate and the mean magnitude of shocks, as well as in relation to the drift rate ($\gamma$) and the volatility ($\sigma$). In this phase of the simulation, we employed contour plots instead of three dimensional surfaces. This methodological choice serves two purposes: first, to provide diversity in the simulation approaches; and second, to demonstrate alternative analytical techniques for visualizing the effects of two interacting parameters on the evolution of trust. Specifically, we explored how trust responds to changes in the drift rate and mean shock magnitude, as well as to the combined influence of drift and volatility both of which are critical drivers of the stochastic behavior modeled through the Lévy process. Naturally, this framework allows for extending the analysis to assess the effects of other parameter combinations on trust dynamics, offering a flexible and robust tool for exploring the multifaceted nature of trust formation and degradation.

To investigate the joint effect of the second player's intrinsic inclination toward trust-building ($\gamma$) and the average magnitude of sudden behavioral shocks ($\mu_J$) on the final level of trust, simulations based on the Lévy process were conducted. For each combination of these two parameters, the model was simulated 20 times, and the mean final trust level was recorded. The results are visualized using a contour plot, which illustrates how the trust level varies across a range of values for $\gamma$ and $\mu_J$.

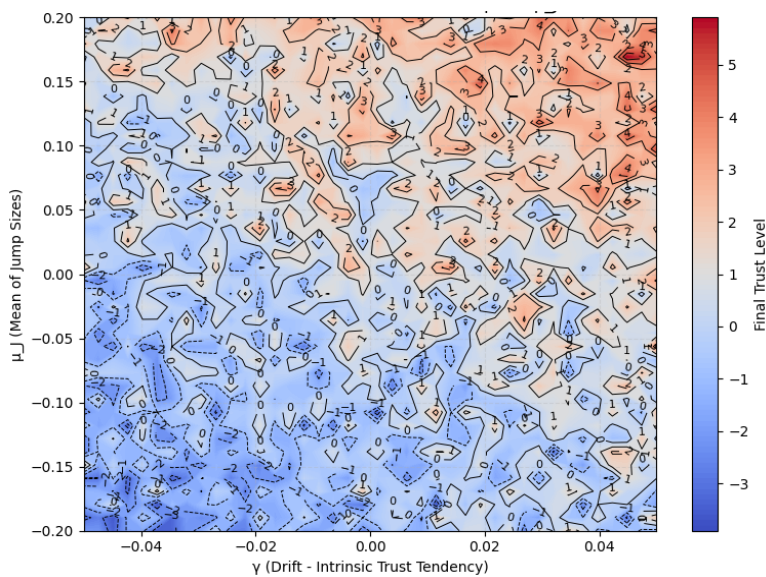

Figure 9.The horizontal axis represents the intrinsic drift rate ($\gamma$), and the vertical axis denotes the average magnitude of behavioral shocks ($\mu_J$). Each colored region indicates the final level of trust after 20 rounds of interaction. Contour lines and numerical labels provide a more precise depiction of trust levels. Warmer colors (red and orange) correspond to higher trust, while cooler colors (blue) indicate lower trust.

*Scientific and Behavioral Analysis of the Contour Plot*

As the value of $\mu_J$ (i.e., the average inclination of shocks toward rewarding and trust enhancing behaviors) increases, the final level of trust rises significantly, even when the overall inclination of the second player ($\gamma$) remains relatively negative. This indicates that positive sudden behaviors can sustain or elevate trust, even under conditions where the second player's general tendency is untrustworthy.

Conversely, in regions where both $\mu_J$ and $\gamma$ are negative, trust levels decline markedly. These areas represent scenarios in which the second player is not only inherently untrustworthy over the long term but also exhibits a propensity for betrayal in sudden behaviors. Under such circumstances, trust rapidly deteriorates.

An additional noteworthy feature is the intermediate regions of the plot where $\gamma$ and $\mu_J$ are both close to zero. Here, trust levels remain moderate and do not exhibit extreme fluctuations. This reflects scenarios characterized by relatively neutral behavior from the second player.

*Contour Plot of Final Trust Level as a Function of Drift Rate and Environmental Volatility*

In this simulation, we investigated the joint effects of the drift rate ($\gamma$) and environmental volatility ($\sigma$) on the final level of trust. The parameter $\gamma$ represents the long-term inclination of the second player toward either trustworthiness or betrayal; positive values indicate a tendency toward being trustworthy, while negative values suggest a propensity for betrayal. The parameter σ captures the intensity of environmental fluctuations; higher values correspond to more unstable and risk-prone interactions.

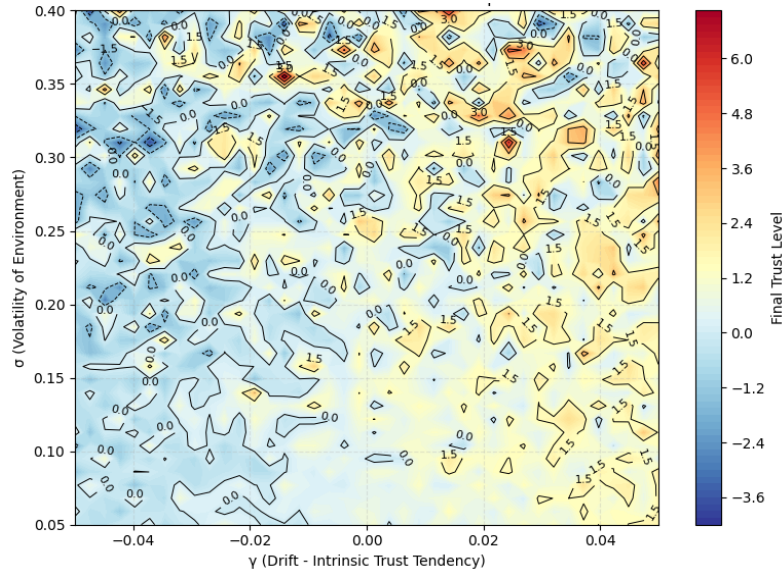

Figure 10.The contour plot illustrates variations in the final level of trust in response to simultaneous changes in the drift rate ($\gamma$) and environmental volatility ($\sigma$). Warmer colors indicate higher levels of trust at the end of the game (with redder regions representing stronger trust and bluer areas reflecting diminished trust).

Contrary to expectations, this plot reveals that the final level of trust is not a monotonic function of either $\gamma$ or $\sigma$ alone; rather, it exhibits highly variable regions with nonlinear patterns. In particular, in areas where $\gamma \approx 0$ and $\sigma$ is high representing a neutral disposition from Player 2 but a highly volatile environment final trust levels may sharply decline. Conversely, at certain points with positive $\gamma$ and moderate or high $\sigma$, trust levels unexpectedly increase, suggesting that inherently positive behavioral tendencies can have a reinforcing effect even in unstable contexts.

This analysis demonstrates that the impact of Lévy model parameters on final trust is nonlinear, dependent on the parametric interplay, and highly sensitive to stochastic conditions. This stands in contrast to classical trust models, which typically display monotonic and deterministic behavior, and highlights the superior explanatory power of our model in capturing real-world behavioral and environmental instabilities.

**Conclusion**

In this paper, we proposed a novel framework for modeling the dynamics of trust between interacting agents based on Lévy jump processes. Unlike classical trust game models that assume either discrete decisions or continuous and smooth changes in behavior, our model incorporates sudden jumps in trust levels allowing it to capture a more realistic range of behaviors, such as unexpected betrayal or spontaneous acts of generosity. These jumps were mathematically modeled by combining Brownian motion with random jumps governed by a Poisson process, representing rare but impactful events in human interactions.

Moreover, decision making noise reflecting ambiguous environments, institutional distrust, or incomplete information was captured through a Gaussian component ($\sigma$). This allowed the model to account for minor but persistent fluctuations in trust. The interaction between this continuous noise and the jump driven behaviors generated inherently unstable and unpredictable trust trajectories. As a result, the multilayered randomness in the model including continuous noise, unpredictable shocks, and the internal disposition of Player 2 enabled the simulation of a wide spectrum of psychological, social, and economic scenarios.

Numerical analyses and contour plots demonstrated that the interplay of parameters such as the mean jump size ($\mu_J$), jump intensity ($\lambda$), the internal drift of Player 2 ($\gamma$), and the volatility of the environment ($\sigma$) can affect the final level of trust in complex, nonlinear, and sometimes contradictory ways. This underscores the importance of a multi-causal and non-deterministic approach in analyzing human behavior an aspect often overlooked in traditional trust game models.

Overall, our Lévy based model provides a more dynamic, behaviorally rich, and psychologically realistic representation of the processes underlying the formation and collapse of trust. It holds significant promise for applications in the analysis of social, economic, and even diplomatic systems where decision-making occurs under uncertainty. Furthermore, this framework lays the groundwork for the development of advanced risk assessment tools and policy strategies in contexts where trust is both volatile and critical.

**AI Tools Acknowledgment**

The core idea, theoretical model, and all analyses presented in this paper are entirely the result of the author's own intellectual work. During the preparation and revision of the manuscript, the author used the generative AI tool ChatGPT (OpenAI, GPT 4, May 2024) solely to improve the readability and language of the text, under full human oversight and with careful review of all AI generated suggestions.

**declaration of interest statement**

The author(s) declare no competing interests

Display quotations of over 40 words, or as needed.

- $X_0$ denotes the initial level of trust.
-

(5) $\sigma$: Introduces fluctuations, but its long-term effect is dominated by drift and jump components.

(6)

$$\text{Displayed equation} \qquad ( )$$

***Simulation Scenarios for the Lévy-Based Trust Game Model***

*Trust dynamics influenced by unpredictable behavior of the counterpart*

*Behaviora Behavioral Dynamics of Players in the Lévy-Process-Based Trust Game under the Scenario of Unstable Trust with Scattered Shocks*

Acknowledgements, avoiding identifying any of the authors prior to peer review

1. This is a note. The style name is Footnotes, but it can also be applied to endnotes.

References: see the journal's instructions for authors for details on style

Table 1.Parameter configuration of the Lévy-based model for simulating erratic and unstable behavioral dynamics. This setup captures a scenario in which trust is subject to intense fluctuations and alternating behavioral shocks, with no clear upward or downward trend—mirroring real-world conditions of unpredictability and instability in interpersonal interactions.

Figure 1.The contour plot illustrates variations in the final level of trust in response to simultaneous changes in the drift rate (γ) and environmental volatility (σ). Warmer colors indicate higher levels of trust at the end of the game (with redder regions representing stronger trust and bluer areas reflecting diminished trust).